\documentclass [preprint,aps]{revtex4}
\usepackage{graphicx}
\def\be{\begin{eqnarray}&&}

\def\psla{\slash \! \! \!}

 \def\ee{\end{eqnarray}}
 \begin{document}
\title{Elastic electron-deuteron scattering and two-body current operators in the 
  Light-Front Hamiltonian Dynamics}
\author{Tobias Frederico $^a$, J. Adnei Marinho $^b$,
Emanuele Pace $^c$, 
Giovanni Salm\`e $^b$} 
\affiliation{$^a$ITA, 12.228-900, S\~ao Jos\'e dos
Campos, S\~ao  Paulo, Brazil,$^b$Istituto Nazionale di Fisica Nucleare
- Sezione di Roma \\ 
        P.le A. Moro 2, 00185 Rome, Italy,$^c$Universit\`a di Roma  "Tor Vergata" and Istituto Nazionale di Fisica Nucleare
- Sezione di "Tor Vergata", Via della Ricerca Scientifica 1, 00133 Rome, Italy}
\begin{abstract}

The electromagnetic properties of the deuteron are investigated within
a Light-Front Hamiltonian Dynamics framework, with a current operator that
contains both  one-body and  two-body contributions. In this work, we are
considering new   two-body
contributions, with  a dynamical
nature generated within a Yukawa model and a structure    inspired 
by a recent analysis of the   current operator,
that acts on the three-dimensional valence component  and
fulfills the Ward-Takahashi identity.
Preliminary results for the magnetic moment are shown.

\end{abstract}\maketitle
\section{Introduction}

A careful description of the deuteron, retaining only
a finite number of baryonic and mesonic degrees of freedom,
could help to single out new, experimental signatures, related to
the underlying degrees of freedom (see \cite{qcd,brodsky} for recent reviews of the
 QCD 
effects in  the deuteron). In the present context, {\em careful
 description} means an approach that 
is able to fulfill general properties, like the extended Poincar\'e covariance 
(the discrete
symmetries are included),
Hermiticity and current conservation \cite{LPS1}.

Deuteron 
electromagnetic (em) observables offer 
 a valuable  play ground  for testing 
theoretical ideas (see, e.g., \cite{gross}). In particular, the challenges for theorists appear
very stimulating in view of  
the Th. Jefferson Lab. (TJLAB) upgrading to 12 GeV, that should open
 very  intriguing
scenarios,  as described, e.g., in PAC-34 and PAC-35 Reports \cite{pac}.
But, the list of issues  to be coped with   is long, if one would like to address 
non standard effects, e.g.  like the 6-quark 
bag \cite{qcd,brodsky,gross}. In our opinion (see, also \cite{gross}),  efforts should be 
 invested on the analysis of:
i) the 
consistency between dynamics and operatorial structure of the current; ii) 
the strong
interplay between different ingredients,  e.g. between the dynamical  content
 of the
two-body currents and the nucleon form factors; iii) the two-photon exchange
effects, that
 could affect the extraction of the em form factors 
 (though recent estimates
  \cite{twopho}
assign them a minor role, $\sim 1\%$); iv)
further baryonic degrees of   freedom, like  isobar configurations, 
till now
investigated only within a non relativistic approach \cite{IC}.

Our  aim is to include  new dynamical two-body  contributions into the em current, 
in a Light-Front Hamiltonian Dynamics (LFHD) framework (see, e.g., \cite{KP} for
a review of the Relativistic Hamiltonian Dynamics), for describing the deuteron
 em observables, still satisfying the
extended Poincar\'e covariance. This work expands  the investigation carried out
in Refs. \cite{LPS2,LPS3}, where a current operator containing the one-body term and a
two-body contribution, needed for satisfying the Hermiticity, was considered.
In  the present approach,
we add two-body terms with a dynamical nature,  generated by
the presence of an explicit one-pion exchange  (see also \cite{tobias96}).
 These two-body currents  are 
inspired by an exact analysis of the four-dimensional (4D) current corresponding to a
 Yukawa model in ladder approximation, where two fermions interact with a pseudoscalar, massive 
 boson \cite{WTI} (notice that, within such an approximation,  there is no 
 photon-boson coupling. The 4D current corresponding to the field
 theoretical model is
 projected onto the three-dimensional (3D) LF hyperplane, so that one obtains an
  operator, that i) acts on the 3D LF valence component of the  interacting-system state
  and ii) automatically
  fulfills the 
Ward-Takahashi Identity (WTI). Moreover, in Ref. \cite{WTI}, it has been
shown that one can
properly truncate, in the Fock space, the LF current and still be able to satisfy
the correspondingly truncated WTI. In particular, the Fock expansion is ordered in powers
of the interaction. In order to
 improve the  calculations of the em deuteron observables of Refs. 
 \cite{LPS2,LPS3}, in this work we  consider  
the first-order   LF current operator, obtained by applying the approach of
Ref. \cite{WTI} to an interacting Lagrangian  
 ${\cal L}=-ig_{PS}\bar \Psi \gamma_5 \vec \tau\Psi \cdot {\vec \phi}$, where $\Psi$ 
 is the
  fermion field and ${\vec \phi}$  an isovector
 pseudoscalar boson field. 
 
 In Sec. \ref{frame}, the choice of the reference frame and some generalities of our
 approach will be presented. In Sec. \ref{current}, the  current adopted in
 our LFHD approach will be illustrated.
 In Sec. \ref{results}, the preliminary results, with only a part of the two-body
 dynamical contributions,  will be shown for
 the deuteron magnetic moment. In Sec. \ref{conclusion}  summary and 
 perspectives will be presented.
 
\section{  Choosing the  frame }
\label{frame}
 Our theoretical frame is  the LF Hamiltonian
Dynamics  combined with the Bakamjian-Thomas
construction \cite{BT} of the Poincar\'e generators for an interacting system. Within
such a framework, as it is  well
known, a relativistic square mass operator of a two-body system can be immediately
identified with the Schr\" odinger equation for that system \cite{Coester}. This allows one to
formally embed the standard {\em non relativistic} deuteron wave function into a
fully Poincar\'e covariant description. The only particular care is the
treatment of the spin part, where the Melosh rotation operators \cite{Melosh}
are requested
(see, e.g., \cite{LPS3} for details). Therefore, within the previous approach, one  can    
 rigorously fulfill 
 the Poincar\'e covariance, for an interacting system with a finite number of degrees of
 freedom.
Heuristically, one could say that the Relativistic Hamiltonian Dynamics 
with a Bakamjian-Thomas
construction  allow one to implement a description of an interacting system
that falls between the non-relativistic quantum mechanics
and local relativistic field theories.
This simple view is  further strengthened  onto   the LF hyperplane, since in this 
case a 
sharp and clean separation between the center of mass motion and  
the intrinsic one can be straightforwardly achieved.  Let us also notice that the
LF boosts are kinematical transformations and that the spectral condition  
$P^+=P^0+P_z\geq 0$ favors a description with a finite number of degrees of freedom.

The other ingredient for developing our description of the em deuteron form
factors is the reference frame where we calculate the theoretical observables.
If one had the complete and calculable theory at disposal, the choice of the frame should  not represent an issue, given the full
covariance,. Actually one
is
 dealing with an approximate scheme, and therefore the choice of 
 the reference frame, 
where  one can more easily implement the constrains for preserving the
general properties, becomes strategic. Following Ref. \cite{LPS1},
one can show that in  a Breit frame   where the momentum
transfer is longitudinal, i.e.
$\vec q _\perp =0$, 
the symmetry of the physical process can be exploited for 
 reducing the constraints imposed on the current operator by  
 the  extended Poincar\'e covariance to a simple rotational covariance 
around the $\widehat z\equiv \widehat q~$-axis.  Then, in this frame, any operatorial dependence, 
fulfilling the
rotational covariance around the $z-axis$, is allowed in the construction of a current 
operator, that, in particular,  i) depends
parametrically upon the CM momenta and ii) acts on the
intrinsic variables.  For instance, the matrix elements
of the current operator can be approximated by the ones obtained from one-body Dirac and
Pauli currents, and then, since the rotational covariance is safe, it turns out that   
the extended Poincar\'e covariance can be fulfilled by using the proper 
transformations (Lorentz boost, rotations, translations etc.). Moreover, in order to
implement the Hermiticity, one has to add a term that contains the dynamical
generator of the transverse  rotations, obtaining necessarily a two-body
contribution to the current. It should be pointed out that such a contribution
must be distinct from the ones that we are going to discuss in what follows,
since the last ones  contain dynamical effects in a more explicit way.

\section{ 
Projecting the em current onto the Light-Front hyperplane}
\label{current}
In this Section, a brief illustration of the results of Ref. \cite{WTI}, in
particular the ones directly relevant for the present calculations, is given.

In Refs. \cite{sales00,sales01,adnei07,WTI}, by using the projection of the 4D physical 
quantities onto the 3D
LF hyperplane (i.e. $x^+=x^0+x^3=0$),  and  the Quasi-Potential approach for the
Transition Matrix \cite{wolja}, it has been established
a formally exact, one-to-one
correspondence between
i) the 
 4D Bethe-Salpeter amplitude of an interacting system and 
 the 3D LF  {\em valence} component, ii) the
 matrix elements of the 4D em current and of the   LF current,  both 
 fulfilling the proper
Ward-Takahashi identity. The previous two ingredients allow one 
 to keep separate the {\em trivial }
propagation in the global time ($\sum_{1,N} x^+_i$ for a system with $N$
constituents), and to focus on   the dependence upon  the 
 relative-time propagation,  governed by the internal dynamics of the
 system. In particular, within this approach, the valence component is found to be  the
eigensolution of a 3D dynamical equation, with an 
  effective
3D interaction that is  exactly related to the 4D kernel of the BS equation.
The exact correspondence between 3D and 4D quantities is accomplished through
operators that contain the 4D interaction and  cannot be determined in a simple
way. This difficulty can be overcome in a workable way by developing 
  a possible approximation
 scheme, for constructing solutions,  based on 
  the Fock expansion of the relevant
 quantities. If one truncates the Fock space, one can use the corresponding
 truncated Fock basis, in order to expand  quantities, like 
the  effective interaction and the current operator,  onto the LF
 hyperplane. Clearly, by using a truncated space only the kinematical 
 symmetries can be satisfied within a
 field theoretical approach (given the infinite number of degrees of freedom),
  but if one restricts to a Relativistic Hamiltonian Dynamics approach, where a
  finite number of degrees of freedom is taken into account, then one can
  adopt the truncated operators and recover the full Poincar\'e covariance. In
  particular, if the truncated current  contains operators that satisfy  the
  covariance around the $z$-axis, and furthermore one uses the valence wave 
  functions,
  eigensolution of the properly truncated mass equation, in the evaluation of the
  matrix elements, then both the current conservation and the  
  Poincar\'e covariance can be implemented. It should be stressed that the
  truncated mass operator has to properly commute with the  Poincar\'e
  generators, as requested by the Bakamjian-Thomas construction.

In the actual calculations, there is another relevant issue for the LF
projection, that makes  the application of the procedure to the fermionic case 
sharply different from the bosonic one. 
The Dirac propagator can be separated in an 
on-shell term and   an instantaneous  (in LF time !)
 propagation, viz
\be iS(k)=
\frac{\psla{k}+m}{k^2-m^2+i\varepsilon}=
\frac{\psla{k}_{on}+m}{k^+(k^--k^-_{on} +{i\varepsilon\over
k^+})}+\frac{\gamma^+}{2 k^+}  \ee
where $k^-_{on}=(|{\bf k}_{\perp}|^2+m^2)/k^+$ is the minus-component of
$k^\mu_{on}$, such that $k_{on}\cdot k_{on}=m^2$ and the second term is
 the instantaneous  one, as shown by  the  Fourier transform in $k^-$ and 
 ${\bf k}_{\perp}$, i.e.
$$\int dk^- ~d{\bf k}_{\perp}~ exp[-i(k^-x^+/2-i{\bf k}_{\perp}
\cdot {\bf x}_{\perp})] = (2 \pi)^3~
\delta(x^+) \delta({\bf x}_{\perp} )~.$$

The instantaneous term  has a great impact on the analysis
 of the matrix elements of the current operator, and it produces
very peculiar operatorial structures in the many-body contribution to the  LF
current. It should be emphasized that this fact is related to our choice 
of the reference frame, where
$q^+\ne 0$, since in this case the extraction of the em form factors involves matrix elements
of both the plus and perp components of the current.  In the frame with $q^+\ne
0$ the Poincar\'e covariance
can be demonstrated within a LFHD approach \cite{LPS1}, and 
therefore the famous  {\em angular
condition} issue (see, e.g. \cite{ang}) does not any more plague  the matrix
elements.

In Ref.  \cite{WTI}, the explicit
 expressions for  many-body terms  of the em  LF current have been 
 obtained, within a Yukawa model in ladder approximation  for two fermions. In 
 our approach 
  for evaluating   the em
 deuteron observables, we take into
 account the current only up to the first-order in the interaction  
 (see  the next Section), as shown 
 by the  diagrammatic representation  of Fig. 1. It is worth noting that one has at least three particles
in flight (the LF-time flows  from right to left).

\begin{figure}

\vspace{0.4cm}
\parbox{10.5cm}{\includegraphics[width=4.cm] {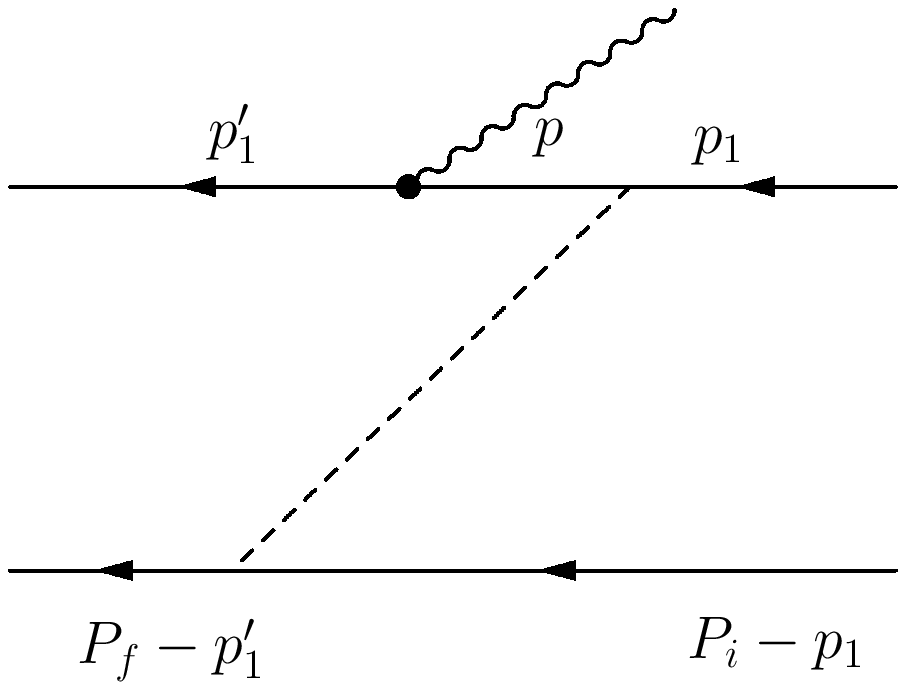}
$\hspace{1.5cm}$\includegraphics[width=4.cm] {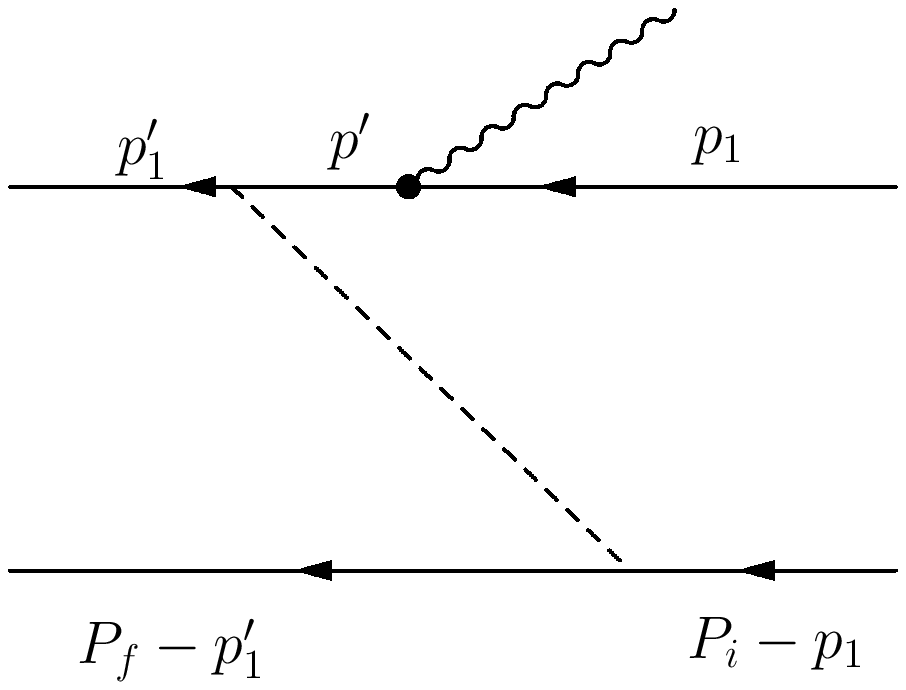}}
$\hspace{1.5cm}$\parbox{2.5cm}{Interaction terms}

\vspace{0.4cm}
\parbox{8cm}{\hspace{3cm}\includegraphics[width=4.cm] {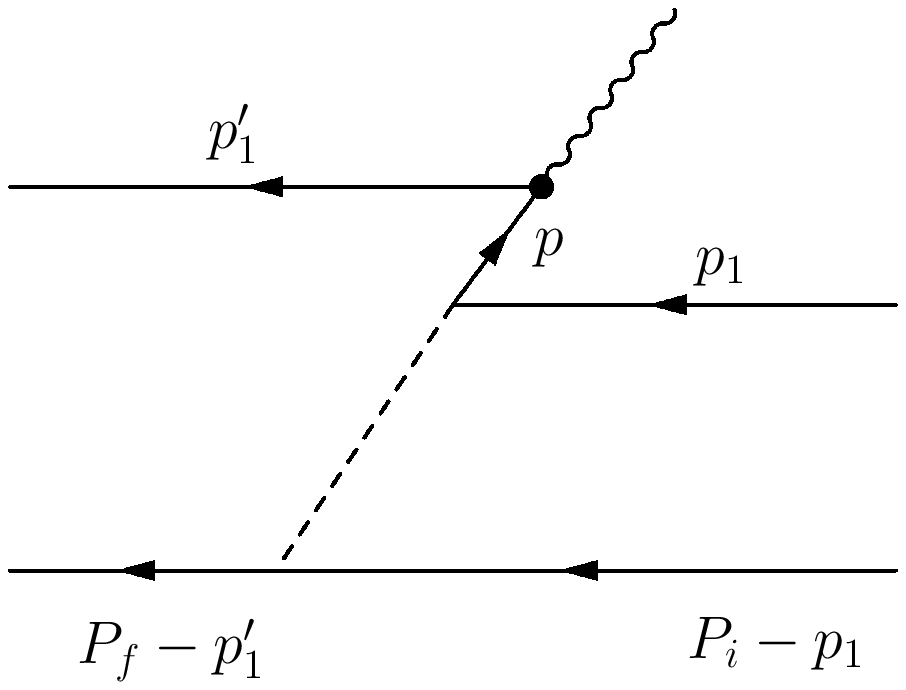}}
$\hspace{4.0cm}$\parbox{2.5cm}{Pair term}

\vspace{0.4cm}
\parbox{10.5cm}{\includegraphics[width=4.cm] {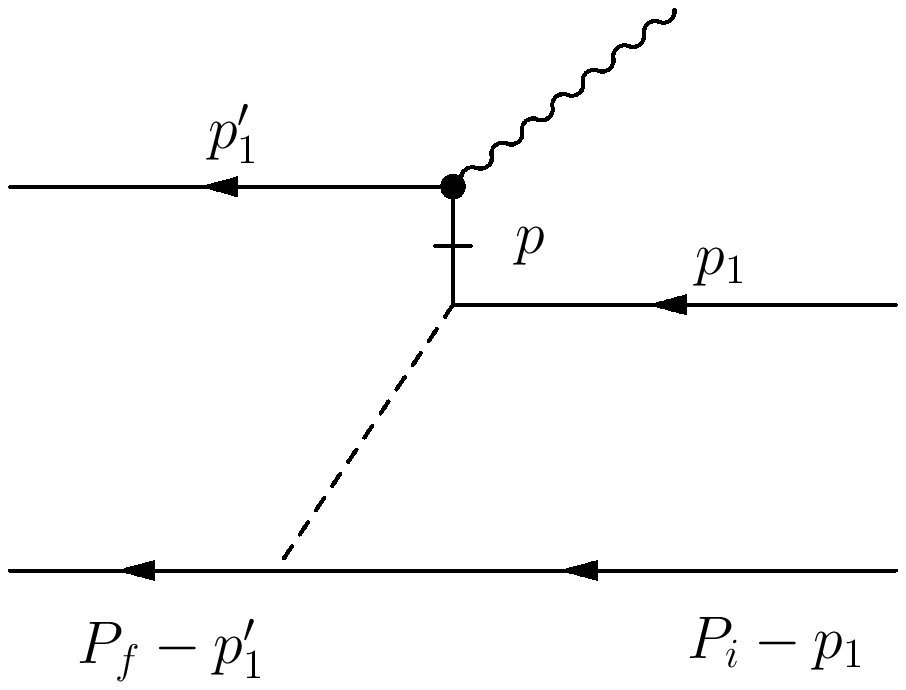}
$\hspace{1.5cm}$\includegraphics[width=4.cm] {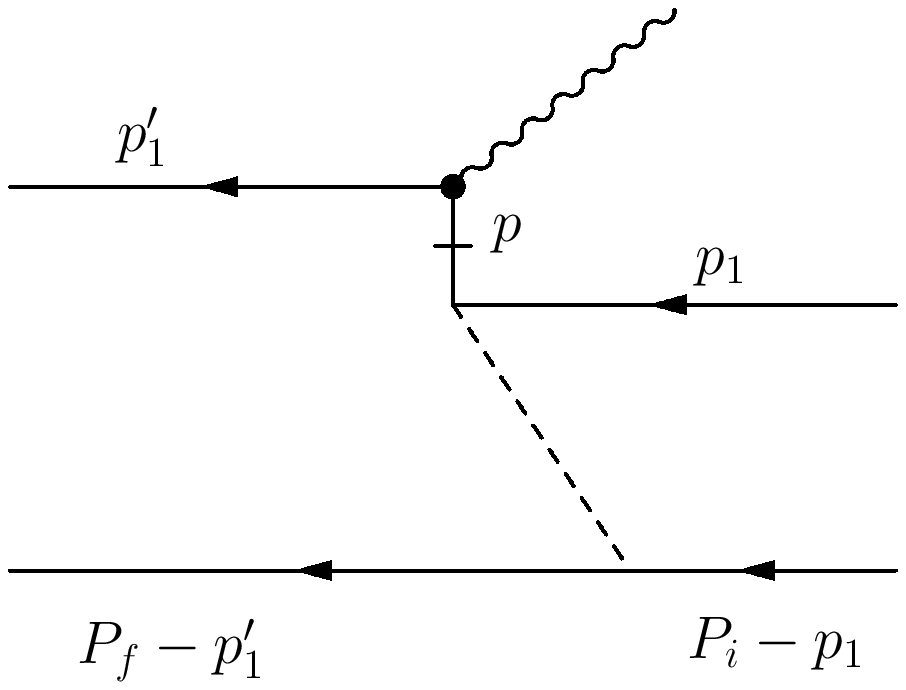}}
$\hspace{1.5cm}$\parbox{2.9cm}{ Instantaneous, in LF time, terms}

\vspace{0.4cm}
\parbox{10.5cm}{\includegraphics[width=4.cm] {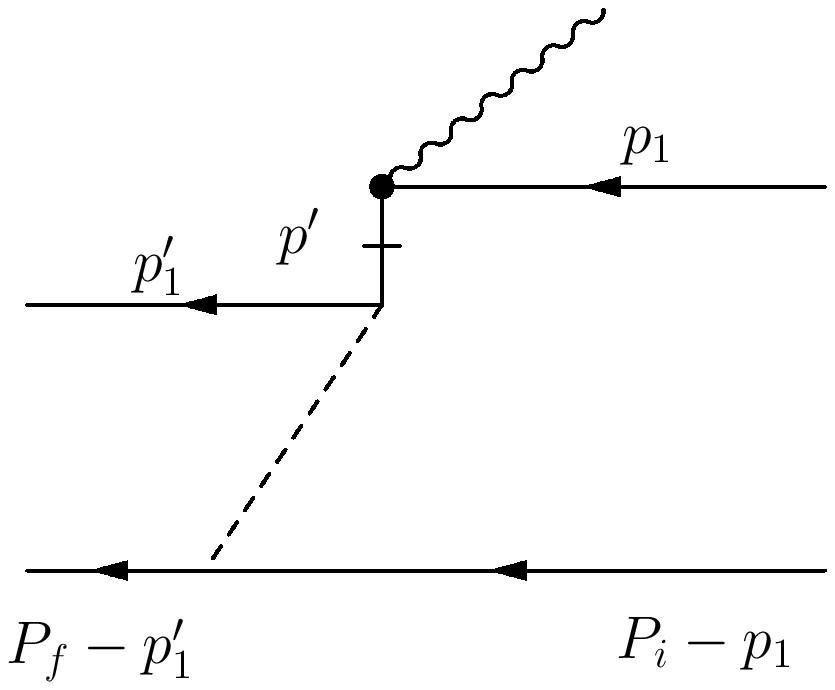}
$\hspace{1.5cm}$ \includegraphics[width=4.cm] {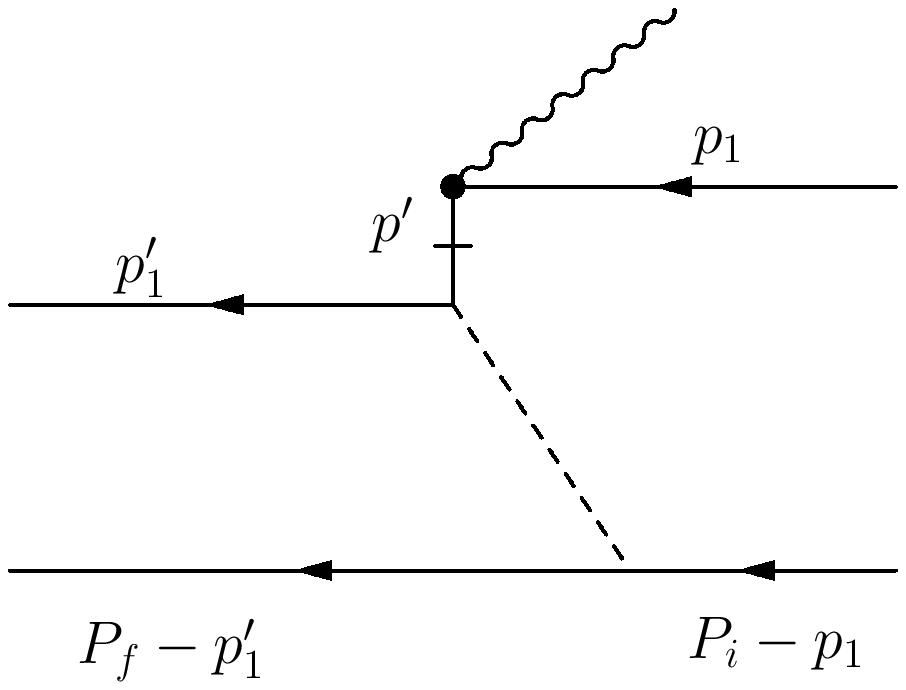}}
$\hspace{1.5cm}$ \parbox{2.9cm}{  Instantaneous, in LF time, terms}

\caption{Diagrammatic analysis of the LF first-order em current, 
obtained within a Yukawa model in
ladder approximation. In Ref. \cite{WTI} one can find 
more details and explicit expressions.}
\end{figure}

\section{Preliminary results for the deuteron magnetic moment }
\label{results}

In a Breit frame where
${\bf q}_\perp=0$ and therefore 
$q^+ \ne 0$,
the matrix elements of the em current operator for an 
interacting system can
be
defined in terms of the   LF current  and the dynamical transverse component of
the Bakamjian-Thomas rotation generator $\vec S_\perp$ 
 as follows
\be  
j^{\mu}(q\hat{e}_z)={{\cal{J}}^{\mu} (q\hat{e}_z)\over 2} +
L^{\mu}_{\nu}[r_x(-\pi)]~e^{\imath \pi S_x}
{{\cal{J}}^{\nu}(q\hat{e}_z)^*\over 2} 
e^{-\imath \pi S_x} \label{cur1} \ee
where $L^{\mu}_{\nu}[r_x(-\pi)]$ is the 4D representation of the $\pi$-rotation
around the $x$-axis, and generates a term necessary for implementing
Hermiticity. In Eq. (\ref{cur1}), the operator ${\cal J}^\mu(q\hat{e}_z) $ is
given by
 \be {\cal J}^\mu(q\hat{e}_z) = 
j_{one}^\mu + j_{two}^\mu=\overline \Pi_0\mathcal{J}^\mu_0\Pi_0+\overline
\Pi_0\left[V\Delta_0\mathcal{J}^\mu_0 +\mathcal{J}^\mu_0 \Delta_0V
\right]\Pi_0
\label{cur2}\ee
where
 $\Pi_0$  is the so called {\em free reverse LF projector} of Ref.
 \cite{WTI}, that singles out the positive-energy sector (modulo some
 kinematical factors)and  $\Delta_0\equiv G_{free}-
G_{glob}$, with   $G_{free}$   the standard two-fermion free propagator and 
$G_{glob}$ an auxiliary Green's function, that takes into account the 
global-time propagation. The operator   $V$ 
is the interaction, mediated by  
a  pion (see also \cite{tobias96})
\be 
V= i~g^2~ \vec\tau_{1}\cdot \vec\tau_{2}~
{\gamma^5_{1}~ \otimes ~\gamma^5_{2}~{\cal F}^2[ (\hat p_2 -\hat p_1 )^2]
\over
\left[(\hat p_2 -\hat p_1 )^2 -m^2_\pi +i \epsilon\right]}\ee
with ${\cal F}[ (\hat p_2 -\hat p_1 )^2]$  the 4D vertex form factor counterpart of the
nonrelativistic vertex function of 
Ref. \cite{CDBON}. The quantities with hats represent proper operators.
In Eq. (\ref{cur2}), the one-body contribution (see
\cite{LPS2,LPS3} for details) is obtained from
\be \mathcal{J}^\mu_0 =\sum_{i=1,2}\left [ J_{p i}^{\mu}(0) {(1+\tau_{3i})\over 2}
+ J_{n i}^{\mu}(0){(1-\tau_{3i}) \over 2}\right]\ee
where the free nucleon current is
\be J^\mu_{N}= -F_{2N}[ (\hat p' -\hat p )^2]{(p^\mu+p^{\prime
\mu})\over 2M} +\gamma^\mu (F_{1N}[ (\hat p' -\hat p )^2]+
F_{2N}[ (\hat p' -\hat p )^2])\ee
with $F_{1N}$ and $F_{2N}$ the Dirac and Pauli form factors, respectively.

What about current conservation and charge 
normalization?
In the chosen Breit frame,  current conservation and charge normalization
read respectively as follows
\be
\langle P_f,d|j^{+}(q\hat{e}_z)|d;P_i\rangle=
\langle P_f,d|j^{-}(q\hat{e}_z)|d;P_i\rangle 
\ee and \be
\hspace{-1 cm}\langle P_i,d|j^{+}(0))|d;P_i\rangle= 
\langle P_i,d|{1 \over 2}\left[{\cal{J}}^{+}(0)+
{\cal{J}}^{-}(0)\right]|d;P_i\rangle
=~e
\ee
If WTI is fulfilled, then one obtains the current conservation, once    matrix elements
are taken between eigensolutions of the mass equation constructed from the proper Green' function. 
This is not
the case in our phenomenological calculations, since we are adopting the 
deuteron wave
functions corresponding to realistic interaction, like CD-Bonn \cite{CDBON} or
AV18 \cite{AV18}. 
But, in the elastic processes like the one we are considering, 
current conservation follows after implementing 
 Hermiticity \cite{LPS1}, given by (note the change of the $z$-axis)
 \be\langle P_f,d|j^{\mu}(q\hat{e}_z)|d;P_i\rangle=
\langle P_i,d|j^{\mu}(-q\hat{e}_z)|d;P_f\rangle^*\ee
Notice that the charge normalization can be fulfilled if in Eq.
(\ref{cur1}) one defines
 \be\langle P_i,d|{\cal{J}}^{-}(0)|d;P_i\rangle=\langle P_i,d|
 {\cal{J}}^{+}(0)|d;P_i\rangle\ee
 This leads to assume  that such an equality holds for any momentum transfer.
It should be pointed out that
 for evaluating  the em observables only  $j^{+}(q\hat{e}_z)$
and $j^{1(2)}(q\hat{e}_z)$ are relevant.

Preliminary results for the magnetic moment of the deuteron
have been obtained by retaining only  the two-body
interaction
terms, corresponding to the diagrams
depicted in the first line
of Fig. 1, and with the explicit expressions given in Ref. \cite{WTI} 
(modulo the isospin dependence and the pseudoscalar coupling). The nucleon
form factors adopted in the calculations are the ones of Ref. \cite{nucleon},
that are in nice agreement with the most recent measurements of the proton form
factors (see, e.g., \cite{pucket}) and represent a first microscopical
interpretation of the possible zero in the ratio $\mu_p G^p_E/ G^p_M$ for $Q^2>8
~(GeV/c)^2$ in terms of interference between the valence and non valence
component of the proton state. It is worth noting that the behavior of the
nucleon form factors, in a wide kinematical region, enters in the evaluation of the
static ($Q^2 \to 0$) em properties of the deuteron, according to the approach of
Refs. \cite{LPS2,LPS3}.
In Table 1, the preliminary values, for different deuteron wave functions,
corresponding to three realistic NN interactions, are shown. In particular,
the CD-Bonn \cite{CDBON}, RSC93 \cite{RSC93} and AV18 \cite{AV18} interactions have been
used. The calculations corresponding to the one-body contribution with the 
Hermiticity
term \cite{LPS2,LPS3} have been also shown.  It should be mentioned that, 
within the last approximation scheme,
 the  experimental quadrupole moment is fairly well described, with an 
 underestimate of the order of $4\%$.
 \begin{center} 
\begin{tabular}{|c|c|c|c|c|}
\hline
Interaction & $P_D$   & $\mu _D^{NR}$ &
$\mu^{LFD}_{one}$   & $\mu^{LFD}_{1+2}$ \\
\hline  
CD-Bonn  & 4.83 &  0.8523  & 0.8670 &  $0.863 \pm 0.002$ \\
RSC93    & 5.70 & 0.8473  & 0.8637 &  $0.861 \pm 0.002$ \\ 
Av18     & 5.76 &  0.8470  & 0.8635& $0.860 \pm 0.002$  \\
\hline 
\end{tabular} 

Exp.  0.857406(1)
\end{center}
The results appear encouraging. But, for a vanishing momentum transfer (see
the analogous discussion for the one-body case in Refs. \cite{LPS2,LPS3}),
it is necessary an accurate 
study of the numerical convergence
of the multifold integrals,  that enter the
calculations.
 From the charge normalization, one can obtain the probabilities of the
valence and non valence components. 
At the present stage we have obtained $Prob_{NV}\sim0.01$.

\section{ Summary \& Perspectives}
\label{conclusion}
In this contribution, we have presented preliminary results
for the magnetic moment of the deuteron,  including 
two-body, dynamical contributions \cite{WTI} to
 em current, within the LFHD approach proposed in Refs. \cite{LPS1,LPS2,LPS3}.
The approach is fully Poincar\'e covariant. 

The new two-body terms have been inspired by an exact analysis of a Yukawa
model, in ladder approximation, for two interacting fermions, carried out in Ref.
\cite{WTI}. It turns out that, 
 onto the LF
hyperplane, one obtains  a LF current fulfilling the Ward-Takahashi identity, 
at
each order in the Fock expansion.
 
The  systematic analysis of the deuteron em form factors is started, 
and it represents a non trivial
task from the numerical point of view, given the many, multidimensional 
integrals
to be performed with very high accuracy, in particular at low $Q^2$.
 First results obtained by using i) the deuteron wave functions corresponding to 
  CD-Bonn, RSC93  and AV18 NN potentials and ii) the 
 interaction
contribution to the two-body current, as depicted in the first line of Fig. 1,
 appear
consistent with the expectations, namely a very low probability for the 
component beyond
the valence one, and a magnetic  moment in fair agreement with the
experimental values. Calculations of the quadrupole moments, more delicate from the numerical point
of view, are in progress.

\section{Acknowledgments}
This work was supported by the Brazilian agencies CNPq and
FAPESP and by the Italian MUR.

\end{document}